\documentclass[%
 reprint,
preprintnumbers,
 amsmath,amssymb,
 aps,
]{revtex4-1}

\usepackage[dvipdfmx]{graphicx,color}
\usepackage{dcolumn}
\usepackage{bm}
\usepackage{hyperref}
\usepackage{xcolor}
\hypersetup{
    colorlinks=true,
    citecolor=blue,
    linkcolor=red,
    urlcolor=violet,
}

\begin{document}

\preprint{KOBE-COSMO-19-21}

\title{Pulsar timing residual induced by ultralight vector dark matter}

\author{Kimihiro Nomura}
 \email{190s111s@stu.kobe-u.ac.jp}
\author{Asuka Ito}%
 \email{asuka-ito@stu.kobe-u.ac.jp}
\author{Jiro Soda}%
 \email{jiro@phys.sci.kobe-u.ac.jp}
\affiliation{%
Department of Physics, Kobe University, Kobe 657-8501, Japan
}%

%
%

\date{\today}

\begin{abstract}
We study the ultralight vector dark matter with a mass around $10^{-23}\,\mathrm{eV}$.
The vector field oscillating coherently on galactic scales induces oscillations of the spacetime metric with a frequency around nHz, which is detectable by pulsar timing arrays. 
We find that the pulsar timing signal due to the vector dark matter has nontrivial 
angular dependence unlike the scalar dark matter and the maximal amplitude is three times 
larger than that of the scalar dark matter.
\end{abstract}

\maketitle


\section{Introduction}

Observations of the galactic rotation curves \cite{Rubin:1982kyu}, structure formation \cite{Davis1985}, and gravitational lensing \cite{Clowe2006} suggest that the invisible matter, the so-called dark matter (DM)
exists in the Universe.
Searching for the DM has been a long-standing challenge in cosmology and astrophysics.
Recent observational results show that the DM is accounting for 27\% of the energy density in the Universe \cite{Aghanim2018}.
The most promising candidate for the DM has been the weak interacting massive particles (WIMPs)
which are  motivated by  supersymmetric theories of particle physics.
However, inspite of the efforts of many researchers, no signal of the WIMPs has been detected.

Recently, as an alternative candidate for the DM, 
an ultralight axion-like scalar with a mass $\sim 10^{-23}\,\mathrm{eV}$, 
often called the fuzzy DM, has been intensively studied~\cite{marsh2015,Hui:2016ltb}.
The fuzzy DM has a possibility to
resolve small-scale problems of the standard Cold DM model such as the galactic core-cusp problem \cite{Hu2000, Press:1989id, Sin:1992bg}.

Given the success of the ultralight scalar DM, it is natural to ask if the ultralight vector can be the DM.   
In fact, the possibility of the fuzzy DM being a massive vector boson (sometimes called a dark photon)
has been proposed~\cite{Nelson2011, Nakayama2019}.
In the case of a vector boson, it is known that it is difficult for a free field to condense homogeneously during inflation.
Recently, however, a consistent ultralight vector model has been 
proposed~\cite{Nakayama2019} where
the ultralight vector field is homogeneously condensed during inflation when its mass is less than the Hubble scale, 
and starts to oscillate coherently at some epoch after inflation.
The coherently oscillating vector field behaves as a non-relativistic matter and can be a candidate for the DM
as well as the scalar field.

Historically, there have been many works  on the vector DM.
Evolution of cosmological perturbations based on the model of ultralight coherent vector DM field has been studied in \cite{Cembranos2016}. There, it is demonstrated that perturbations on the  scales smaller than the de Broglie wave length of the vector field have a specific feature compared to the scalar case. 
However, the magnitude of the feature is far below the sensitivity of present and future detectors.
The approaches to search for the vector DM or to constrain its couplings to particles in the Standard Model have been proposed, e.g. by using ground-based gravitational-wave interferometers \cite{Pierce2018, Guo2019} or taking into account the cosmological plasma effects with a photon \cite{McDermott2019}.
Some phenomenologies on the vector DM are discussed together with its production mechanism \cite{Arias2012, Graham2015, Agrawal2018, Bastero-Gil2018, Co2018, Masaki:2018eut,Dror2018, Ema2019, Nakayama2019}.
The effects of gravitational interaction between the vector DM field and binary pulsar systems on the dynamics of the system have been considered in \cite{LopezNacir2018}.

In this paper, we  investigate the purely gravitational effects of the vector DM on the pulsar timing.
A coherently oscillating vector field  behaves as a non-relativistic pressureless matter on cosmological scales. 
Hence it does not induce the anisotropic expansion of the Universe \cite{Cembranos2012}.
Actually, however, there exists  an oscillating anisotropic pressure on time scales corresponding to the oscillation of the vector field, which induces nontrivial oscillations of the metric.
The frequency is determined by the mass of the DM considered. In the case of $m \sim 10^{-23} \,\mathrm{eV}$, it is on the order of nHz, which 
 is in the range where the Pulsar Timing Arrays (PTAs) are sensitive.
PTAs can detect gravitational effects such as gravitational 
waves \cite{Detweiler:1979wn,Tsuneto:2018tif, sazhin1978opportunities}.
Furthermore, a method for detecting  the axion-like scalar DM using PTAs has been proposed in~\cite{khmelnitsky2013},
 and actually the energy density of the DM is constrained by using observational data~\cite{Porayko2014, Porayko2018, Kato2019}.
The detection method is applicable to the case of the vector DM and it is expected that a specific signal depending on the vector property appears.
This is what we will discuss in this paper.

This paper is organized as follows.
In Section \ref{sec.metricperturbation}, we show the oscillation of the vector DM field induces time-dependent spacetime metric perturbations.
In Section \ref{sec.pulsartiming}, we see the effect of the vector DM 
on the pulsar timing.  We also discuss the detectability of the vector
DM with PTAs.
In Section \ref{sec.conclusion}, we summarize the results.
Appendix \ref{sec.photonredshift} is devoted to derivation of
 the formula for the redshift of photons induced by metric perturbations.

\section{Effects on metric perturbations}
\label{sec.metricperturbation}

In this section, we show that the spacetime metric fluctuates 
due to the presence of the vector DM oscillating coherently on galactic scales.

Since we work on the galactic scales,  the cosmic expansion is negligible.
Hence, we consider the metric 
\begin{align}
ds^{2} &= \eta_{\mu\nu} dx^{\mu} dx^{\nu}
\notag \\
&\quad - 2 \Phi(t, \mathbf{x}) dt^{2} + 2\Psi(t, \mathbf{x}) \delta_{ij} dx^{i} dx^{j}
+ h_{ij}(t, \mathbf{x}) dx^{i} dx^{j},
\label{eq.vdm.0}
\end{align} 
where $\eta_{\mu\nu}$ is the Minkowski metric $\eta_{\mu\nu} = \mathrm{diag}(-1,+1,+1,+1)$. 
The metric perturbations induced by the DM are described by $\Phi$, $\Psi$, and $h_{ij}$.
We shall discuss the perturbations separately in eq.\,\eqref{eq1.9} and eq.\,\eqref{eq1.16}.

The vector DM field $A_{\mu}$ with a mass $m$ is expected 
to have little interaction with particles in the Standard Model.
We treat the vector DM field as a free field. Its action is given by
\begin{align}
S = \int d^{4}x \sqrt{-g} \left(- \frac{1}{4} g^{\mu \nu} g^{\rho \sigma} F_{\mu \rho} F_{\nu \sigma} - \frac{1}{2} m^{2} g^{\mu \nu} A_{\mu} A_{\nu} \right),
\label{eq.vdm.1}
\end{align}
where $g$ is the determinant of the metric $g_{\mu\nu}$ and
$F_{\mu\nu} = \partial_{\mu} A_{\nu} - \partial_{\nu} A_{\mu}$.
We also consider the ultralight mass of the vector DM as $m \sim 10^{-23} \, \mathrm{eV}$, which can arise through the Higgs or the Stueckelberg mechanisms.

Here, we follow the discussion in \cite{khmelnitsky2013}, where  the case of the scalar DM is considered.
Taking into account that the typical velocity in the galaxy is given by $10^{-3}$ times the speed of light, 
 we can estimate the occupation number of the vector DM  as
\begin{align}
&\frac{\rho}{m \cdot (mv)^{3}}
\notag \\
&\quad\simeq 10^{95} \left( \frac{\rho}{0.4 \, \mathrm{GeV}/\mathrm{cm}^{3}} \right) \left( \frac{10^{-23}\, \mathrm{eV}}{m} \right)^{4} \left( \frac{10^{-3}}{v} \right)^{3},
\label{eq0.2}
\end{align}
where the energy density of the DM $\rho$ is normalized 
by the local value~\cite{Catena2010, Salucci2010, Weber2009}.
Since the occupation number is so huge,
 we can treat the vector field as a classical wave.
 The de Broglie wavelength for the ultralight vector DM particles with a mass $m$ reads
\begin{align}
\lambda_{\mathrm{dB}} = \frac{2\pi}{mv} 
\simeq 4 \, \mathrm{kpc} \left( \frac{10^{-23} \, \mathrm{eV}}{m} \right) \left( \frac{10^{-3}}{v} \right)\ .
\label{eq0.1}
\end{align}
 Due to the wave nature of the DM field,  all inhomogeneities in the DM distribution on the scale smaller than $\lambda_{\text{dB}}$ are smoothed out.
So we can express the vector DM field as a superposition of plane waves with a typical wave number $k = mv = 2\pi / \lambda_{\mathrm{dB}}$.
In the present case, the energy reads $E \simeq m + mv^{2}/2\simeq m$, because the velocity is non-relativistic. 
Thus,  the vector DM field is coherently oscillating with a monotonic frequency determined by the mass $m$ on the scale given by $\lambda_{\text{dB}}$.

On galactic scales, we can neglect the cosmic expansion.
The equations of motion of the vector field in a flat background read
\begin{align}
\partial_{\mu} F^{\mu\nu} - m^{2} A^{\nu} = 0,
\label{eq.vdm.4}
\end{align}
which is derived by taking the variation of the action \eqref{eq.vdm.1} with respect to $A_{\mu}$.
Using the identity $\partial_{\nu}\partial_{\mu} F^{\mu\nu} = 0$, we can rewrite eq.\,(5) as a set of Proca equations in the standard form:
\begin{align}
\partial_{\mu} A^{\mu} = 0 \,,
\label{eq.vdm.5-re}\\
(- \partial_{0}^{2} + \bm{\nabla}^{2} - m^{2}) A^{\mu} = 0 \,.
\label{eq.vdm.6-re}
\end{align} 
Taking into account that the field typically has a frequency $m$ and a momentum $k$, eq.\,(6) gives
\begin{align}
A_{t} \sim \frac{k}{m} A_{i} \,.
\label{eq.vdm.8-re}
\end{align}
Hence, $A_{t}$ gets  suppressed by the order of $k/m = v \sim 10^{-3}$ compared to $A_{i}$. Thus, we can neglect $A_{t}$.
During inflation, only the longitudinal mode survives. Hence, the directions of the vector at different points in a coherent region align. 
We take a coordinate system so that the direction of the oscillation is along $z$-axis. 
From eq. \eqref{eq.vdm.6-re},  we obtain
\begin{align}
&A_{z}(t, \mathbf{x}) = A(\mathbf{x}) \cos (mt + \alpha(\mathbf{x}))
\label{eq1.1} \ .
\end{align}
Here we neglected the spatial derivative when we solve the equation.
However,  we left spatial dependence of the amplitude and phase.
Note that the scale of variation of these quantities 
is larger than $\lambda_{\mathrm{dB}}$.

Let us see metric perturbations induced by the oscillating DM.
The energy-momentum tensor for matter fields is defined by
\begin{align}
T_{\mu\nu} =  \frac{- 2}{\sqrt{- g}} \frac{\delta S}{\delta g^{\mu\nu}}.
\label{eq.vdm.15}
\end{align}
For a free massive vector field \eqref{eq.vdm.1}, we have
\begin{align}
T_{\mu\nu} &= g_{\mu\nu} \left( - \frac{1}{4} g^{\rho\alpha} g^{\sigma\beta} F_{\alpha\beta} F_{\rho\sigma} - \frac{1}{2} m^{2} g^{\rho\sigma} A_{\rho} A_{\sigma} \right) \notag \\
&\quad + g^{\rho\sigma} F_{\mu\rho} F_{\nu\sigma} + m^{2} A_{\mu} A_{\nu}.
\label{eq.vdm.16}
\end{align}
In a flat background, the components of the energy-momentum tensor 
\eqref{eq.vdm.16} read
\begin{align}
&
T_{tt}
	= \frac{1}{2} m^{2} A^{2}(\mathbf{x}),
\label{eq1.3} \\
&T_{xx} = T_{yy} 
	= - \frac{1}{2} m^{2} A^{2}(\mathbf{x}) \cos (2mt + 2\alpha(\mathbf{x})),
\label{eq1.5} \\
&T_{zz} 
	= \frac{1}{2} m^{2} A^{2}(\mathbf{x}) \cos (2mt + 2\alpha(\mathbf{x}))
	,
\label{eq1.4}
\end{align}
where we have neglected the spatial derivative of the field.
Notice that the energy density of the DM $T_{tt}$ is time-independent.
On the other hand, the anisotropic pressure is time-dependent.
When we average the pressure over cosmological time scales which are much
longer than the oscillation period, the pressure vanishes.
This tells us that a coherently oscillating massive vector field behaves as a non-relativistic matter with zero-pressure on cosmological scales.
As we will see below, however, the oscillating pressure affects
 the spacetime metric on the time scale relevant to the PTAs.

In general, a symmetric $3 \times 3$ tensor $T_{ij}$ can be decomposed into a trace part and a traceless part as
\begin{align}
T_{ij} = \frac{1}{3} \delta_{ij} {T^{k}}_{k} + \left( T_{ij} - \frac{1}{3} \delta_{ij} {T^{k}}_{k} \right) .
\label{eq1.6}
\end{align}
The first term corresponds the trace part, which behaves as a scalar under three-dimensional rotations.
From eqs. \eqref{eq1.5} and \eqref{eq1.4}, we get
\begin{align}
{T^{k}}_{k} = - \frac{1}{2} m^{2} A^{2}(\mathbf{x}) \cos (2mt + 2 \alpha(\mathbf{x})).
\label{eq1.7}
\end{align}
The second term in \eqref{eq1.6}, the traceless part, is
\begin{align}
&T_{ij} - \frac{1}{3} \delta_{ij} {T^{k}}_{k} \notag \\
&\quad = - \frac{1}{3} m^{2} A^{2}(\mathbf{x}) \cos (2mt + 2 \alpha(\mathbf{x}))
	\begin{pmatrix}
			1	&0	&0	\\
			0	&1	&0	\\
			0	&0	&-2
	\end{pmatrix} \ .
\label{eq1.8}
\end{align}
In the rest of this section, we will consider 
the trace part and the traceless part of the energy-momentum tensor separately.
 
First, let us focus on the trace part.
We take the Newtonian gauge and write the perturbed metric as
\begin{align}
ds^{2} = - (1 + 2 \Phi(t, \mathbf{x})) dt^{2} + (1 + 2\Psi(t, \mathbf{x})) \delta_{ij} dx^{i} dx^{j}  \ ,
\label{eq1.9}
\end{align}
where $\Phi$ and $\Psi$  correspond to the gravitational potential.
We can expect that time dependence of the potential is induced by coherent oscillations of the DM field.
For convenience, we write the potential as the sum of a time-independent part and an oscillating part with a frequency $2m$:
\begin{align}
\Phi(t, \mathbf{x}) &= \Phi_{0}(\mathbf{x}) + \Phi_{osc}(\mathbf{x})\cos(2mt + 2 \alpha(\mathbf{x})), 
\label{eq1.9.1} \\
\Psi(t, \mathbf{x}) &= \Psi_{0}(\mathbf{x}) + \Psi_{osc}(\mathbf{x})\cos(2mt + 2 \alpha(\mathbf{x})).
\label{eq1.9.2}
\end{align}
Given the energy-momentum tensor, the $tt$ component of linearized Einstein equations is
\begin{align}
\partial_{i}^{2} \Psi = - 4 \pi G T_{tt}.
\label{eq1.9.3}
\end{align}
From the result \eqref{eq1.3}, the right-hand side does not depend on time, so this relation determines the time-independent part $\Psi_{0}(\mathbf{x})$.
In order to find time dependence of the gravitational potential, we use the trace of the spatial component of Einstein equations,
\begin{align}
-3 \ddot{\Psi} + \partial_{i}^{2}(\Phi + \Psi) = 4 \pi G {T^{k}}_{k}.
\label{eq1.10}
\end{align}
Now we substitute eqs. \eqref{eq1.7}, \eqref{eq1.9.1} and \eqref{eq1.9.2} into this expression and split terms into time-dependent and time-independent terms.
Focusing on time-independent terms, we can see $\partial_{i}^{2}(\Phi_{0} + \Psi_{0}) = 0$, which implies $\Phi_{0} = - \Psi_{0}$.
On the other hand, as for the time-dependent parts, 
one can neglect $\partial_{i}^{2}(\Phi + \Psi)$ term 
because the spatial gradients on $\Phi$ or $\Psi$ typically bring out $k$.
Thus these terms are suppressed compared with $\ddot{\Psi}$ term because $k^{2}/m^{2} \sim v^{2}$ is tiny.
Assuming the amplitude of the oscillating part of the gravitational potential, $\Psi_{osc}$, is sufficiently homogeneous over the length scale considered, 
we have 
\begin{align}
\Psi_{osc}(\mathbf{x}) &= - \frac{1}{6} \pi G A^{2}(\mathbf{x})
	= -\frac{\pi G \rho(\mathbf{x})}{3 m^{2}}
	\notag \\
	&= -2.2 \times 10^{-16} \left( \frac{\rho(\mathbf{x})}{0.4 \, \mathrm{GeV}/\mathrm{cm}^{3}} \right) \left( \frac{10^{-23}\,\mathrm{eV}}{m} \right)^{2}
	.
\label{eq1.13}
\end{align}
In the second equality, we used the energy density of the DM $\rho$ 
given by the $tt$ component of the energy-momentum tensor \eqref{eq1.3}.
Moreover, the oscillation frequency is given by
\begin{align}
f = \frac{2m}{2\pi} = 4.8 \times 10^{-9} \,\text{Hz} \left( \frac{m}{10^{-23} \,\text{eV}} \right).
\label{eq1.15}
\end{align}

Next, we consider the effect of the traceless part of the energy-momentum tensor.
We denote traceless metric perturbation as $h_{ij}$:
\begin{align}
ds^{2} &= - dt^{2} + (\delta_{ij} + h_{ij}(t, \mathbf{x})) dx^{i} dx^{j}, \label{eq1.16} \\
{h^{i}}_{i} &= 0.
\notag
\end{align}
In the linearized Einstein equations, a combination $\ddot{h}_{ij} - \partial_{k}^{2} h_{ij}$ appears.
By the same reasoning as the above discussion for the trace part, the contribution from $\partial_{k}^{2} h_{ij}$ term can be neglected compared to $\ddot{h}_{ij}$.
Thus we have
\begin{align}
\ddot{h}_{ij} = 16 \pi G \left( T_{ij} - \frac{1}{3} \delta_{ij} {T^{k}}_{k} \right).
\label{eq1.18}
\end{align}
By using \eqref{eq1.8}, the traceless part of the metric perturbation can be obtained:
\begin{align}
{h}_{ij}(t, \mathbf{x}) &= \frac{4}{3} \pi G A^{2}(\mathbf{x}) \cos (2mt + 2 \alpha(\mathbf{x}))
	\begin{pmatrix}
			1	&0	&0	\\
			0	&1	&0	\\
			0	&0	&-2
	\end{pmatrix}.
\label{eq1.19}
\end{align}
For later convenience, we define the amplitude of the oscillation as
\begin{align}
h_{osc}(\mathbf{x}) &\equiv \frac{4}{3} \pi G A^{2}(\mathbf{x})
=\frac{8 \pi G \rho(\mathbf{x})}{3 m^{2}}\notag \\
&= 1.7 \times 10^{-15}  \left( \frac{\rho(\mathbf{x})}{0.4 \, \mathrm{GeV}/\mathrm{cm}^{3}} \right) \left( \frac{10^{-23}\,\mathrm{eV}}{m} \right)^{2}.
\label{eq1.19.1}
\end{align}
From eq. (\ref{eq1.19}), we find that anisotropic metric perturbations appear.
This effect comes from the anisotropy of the vector DM field.
In the case of the scalar DM such as axion-like particles, this kind of anisotropy does not occur.
Therefore, it is possible to distinguish whether the DM is scalar or not from the presence or absence of anisotropy of the metric perturbations.
Notice that the frequency given by \eqref{eq1.15} is the sensitive region of PTAs.
In the next section, we will evaluate the pulsar timing signals from the vector DM discussed above.

\section{Effects on pulsar timing arrays}
\label{sec.pulsartiming}

First of all, we would like to clarify the picture we envisage.
During inflation, the vector field was frozen in a coherent direction, the vector field started oscillating at some point and acting as a dark matter.
It is legitimate to assume that the coherence of the vector field over the region we are observing survives even in the present universe as is always assumed in the axion dark matter.
In fact, the de Broglie wavelength of the vector dark matter is several kpc,
within which the direction of the vector field is coherent.
In the Milky Way, there are many domains of the vector condensations.
The direction of a domain is different from another domain.
We assumed that we are in one of them where there are many pulsars.
The aim of this work is to determine the direction of the vector field in our vicinity with the pulsar timing arrays.

Let us investigate how time-dependent metric perturbations due to the coherently oscillating vector DM field affect
the observed periodic electromagnetic fields from pulsars.
We choose a coordinate system so that the observation point is at the spatial origin, and the direction of the vector DM oscillation is along $z$-axis
as is done in \eqref{eq1.1}. 
A unit vector pointing from the observer to a pulsar is written as
\begin{align}
\mathbf{n} = (\sin \theta \cos \phi, \sin \theta \sin \phi, \cos \theta).
\label{eq1.20}
\end{align}
Then the pulsar is located at $\mathbf{x}_{p}(=|\mathbf{x}_{p}| \mathbf{n})$.
The rotational period of the pulsar is 
$T_{0} = 2\pi/\omega_{0}$, where we have introduced the angular frequency $\omega_{0}$.
Moreover, the observed angular frequency of the pulses is denoted
by $\omega_{obs}(t)$.
Then, the redshift of electromagnetic fields propagating from the pulsar to the observer
is defined by
\begin{align}
z(t) \equiv \frac{\omega_{0} - \omega_{obs}(t)}{\omega_{0}} \ .
\label{eq1.21}
\end{align}
Since we have a relation
 $z(t) =  - ( \omega_{obs}(t) - \omega_{0} ) / \omega_{0}
 \simeq ( T_{obs}(t) - T_{0} ) / T_{0}$, where $T_{obs}(t) \equiv 2\pi/\omega_{obs}(t) $,
 $z(t)$ stands for the relative variation of the observed pulsar timing.
Then, conventionally, the timing residual with respect to a reference time $t = 0$ 
is defined as
\begin{align}
R(t) = \int_{0}^{t} dt' z(t') \ .
\label{eq1.22}
\end{align}

First, we focus on the effect of scalar perturbations on the pulsar timing, in particular $\Psi$, which is induced by the trace part of the energy-momentum tensor of the DM field.
In this case, the redshift is given by (see Appendix. \ref{aps})
\begin{align}
z_{\Psi}(t) =  \Psi(t, \mathbf{0}) - \Psi(t - |\mathbf{x}_{p}|, \mathbf{x}_{p}).
\label{eq3.4}
\end{align}
Paying attention only to the oscillating part, which is measurable with PTAs, we have
\begin{align}
&z_{\Psi}(t) \notag \\
&= \Psi_{osc} \left[ \cos (2mt + 2\alpha(\mathbf{0})) - \cos(2mt - 2m|\mathbf{x}_{p}| + 2 \alpha(\mathbf{x}_{p})) \right] \notag \\
&= -2 \Psi_{osc} \sin(m|\mathbf{x}_{p}| + \alpha(\mathbf{0}) - \alpha(\mathbf{x}_{p})) \notag \\
&\qquad\times\sin (2mt - m|\mathbf{x}_{p}| + \alpha(\mathbf{0}) + \alpha(\mathbf{x}_{p}) ),
\label{eq3.5}
\end{align}
where we assumed that the oscillation amplitude at the observation point and that at the pulsar are approximately equal, and used the same symbol $\Psi_{osc}$ for them.
Integrating $z(t)$ over time according to \eqref{eq1.22}, 
we can evaluate the timing residual induced by the trace part of the metric perturbations as
\begin{align}
R_{\Psi} &= \frac{\Psi_{osc}}{m} \sin(m|\mathbf{x}_{p}| + \alpha(\mathbf{0}) - \alpha(\mathbf{x}_{p}))\notag \\
&\qquad\times\cos (2mt - m|\mathbf{x}_{p}| + \alpha(\mathbf{0}) + \alpha(\mathbf{x}_{p}) ).
\label{eq3.5.1}
\end{align}
It depends on the distance to the pulsar.

Next, we consider the effect of the  traceless part of metric perturbations $h_{ij}$ on the pulsar timing,  which is induced by the traceless part of the energy-momentum tensor of the DM field.
For this part, the redshift is expressed as (see Appendix. \ref{apt})
\begin{align}
z_{h}(t) = \frac{1}{2} n^{i} n^{j} \left[  h_{ij}(t, \mathbf{0}) - h_{ij}(t - |\mathbf{x}_{p}|, \mathbf{x}_{p}) \right] .
\label{eq3.6}
\end{align}
Substituting the result \eqref{eq1.19} in the previous section  and the definition \eqref{eq1.20} into this expression, we obtain
\begin{align}
&z_{h}(t) \notag \\
&= - \frac{1}{4} (1 + 3 \cos 2 \theta) h_{osc} \notag \\
&\quad \times \left[ \cos (2mt + 2 \alpha(\mathbf{0})) - \cos(2mt - 2m|\mathbf{x}_{p}| + 2\alpha(\mathbf{x}_{p})) \right] \notag \\
&= \frac{1}{2} (1 + 3 \cos 2 \theta) h_{osc} \sin(m|\mathbf{x}_{p}| + \alpha(\mathbf{0}) - \alpha(\mathbf{x}_{p})) 
\notag \\
&\quad \times\sin (2mt - m|\mathbf{x}_{p}| + \alpha(\mathbf{0}) + \alpha(\mathbf{x}_{p}) ).
\label{eq3.7}
\end{align}
Correspondingly, the timing residual reads
\begin{align}
R_{h} &= -\frac{1}{4} (1 + 3 \cos 2 \theta) \frac{h_{osc}}{m} \sin(m|\mathbf{x}_{p}| + \alpha(\mathbf{0}) - \alpha(\mathbf{x}_{p})) \notag \\
&\qquad\times\cos (2mt - m|\mathbf{x}_{p}| + \alpha(\mathbf{0}) + \alpha(\mathbf{x}_{p}) ).
\label{eq3.7.1}
\end{align}
Thus, the timing residual depends on the distance to the pulsar and its angular position with respect to the direction of the vector DM oscillation.

\begin{figure}[t]
\centering
  \includegraphics[width=25mm]{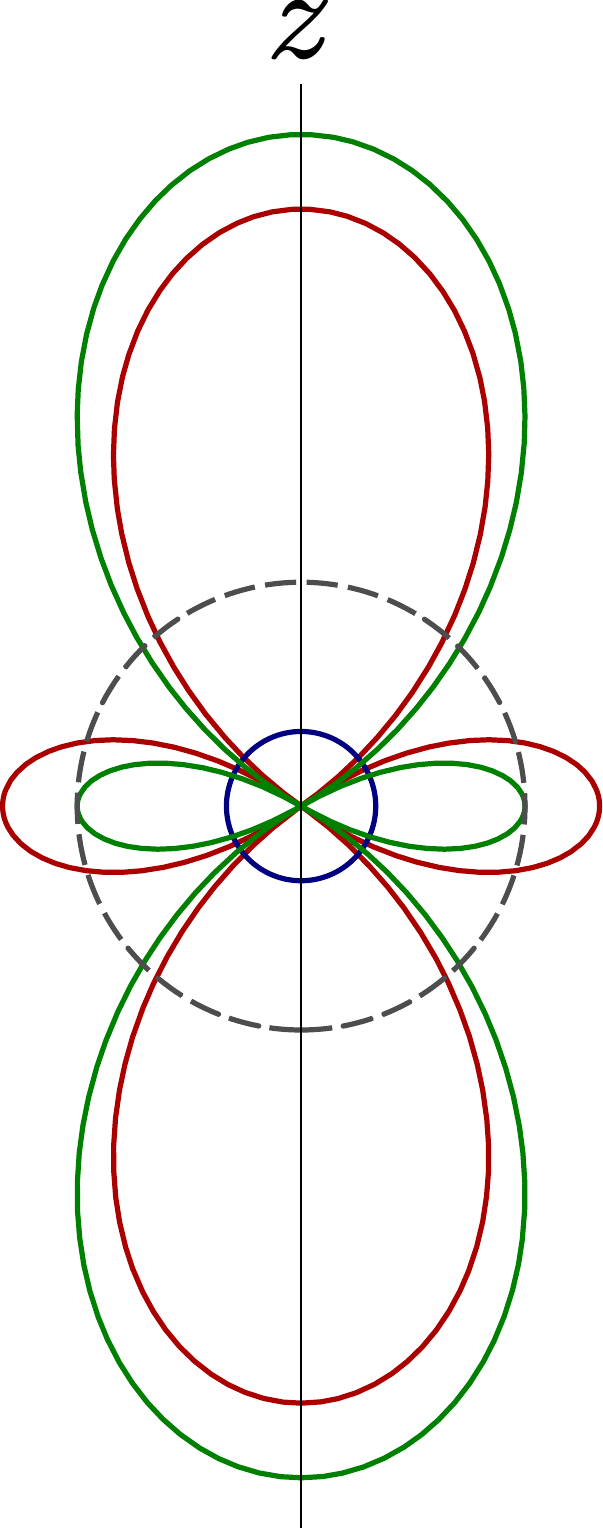}
  \caption{Angular dependence of the redshift due to the oscillation
  of the vector DM  is shown.  The blue line and red line represent the contribution of the trace part $z_{\Psi}$ \eqref{eq3.5}  and the traceless part $z_{h}$ \eqref{eq3.7}, respectively. Actually, we only observe the summation of $z_{\Psi}$ and $z_{h}$, which is  depected by the green line. 
The angle $\theta$ is measured from the direction of 
the oscillation chosen as the $z$-axis.
A gray dashed line shows the magnitude of the redshift when the DM is a scalar field.}
  \label{fig.VectorDM1}
\end{figure}

Recall that the timing residual due to the coherent oscillation of an ultralight scalar DM is~\cite{khmelnitsky2013},
\begin{align}
R_{\mathrm{scalar}} &= \frac{\pi G \rho}{m^{3}} \sin(m|\mathbf{x}_{p}| + \alpha(\mathbf{0}) - \alpha(\mathbf{x}_{p}))
\notag \\
&\qquad\times\cos (2mt - m|\mathbf{x}_{p}| + \alpha(\mathbf{0}) + \alpha(\mathbf{x}_{p}) ).
\label{eq3.7.2}
\end{align}
The main difference is that the timing residual of the vector DM has a nontrivial direction dependence as shown in Fig. \ref{fig.VectorDM1},
in contrast to that of the scalar DM.
Thus, it is possible to discriminate the vector DM from the scalar DM.
Moreover, from eqs. \eqref{eq1.13}, \eqref{eq1.19.1}, \eqref{eq3.5.1}
and \eqref{eq3.7.1}, we can see that the magnitude of the maximal timing residual of the vector DM is three times larger than that of the scalar DM.

\begin{figure}[t]
\centering
  \includegraphics[width=85mm]{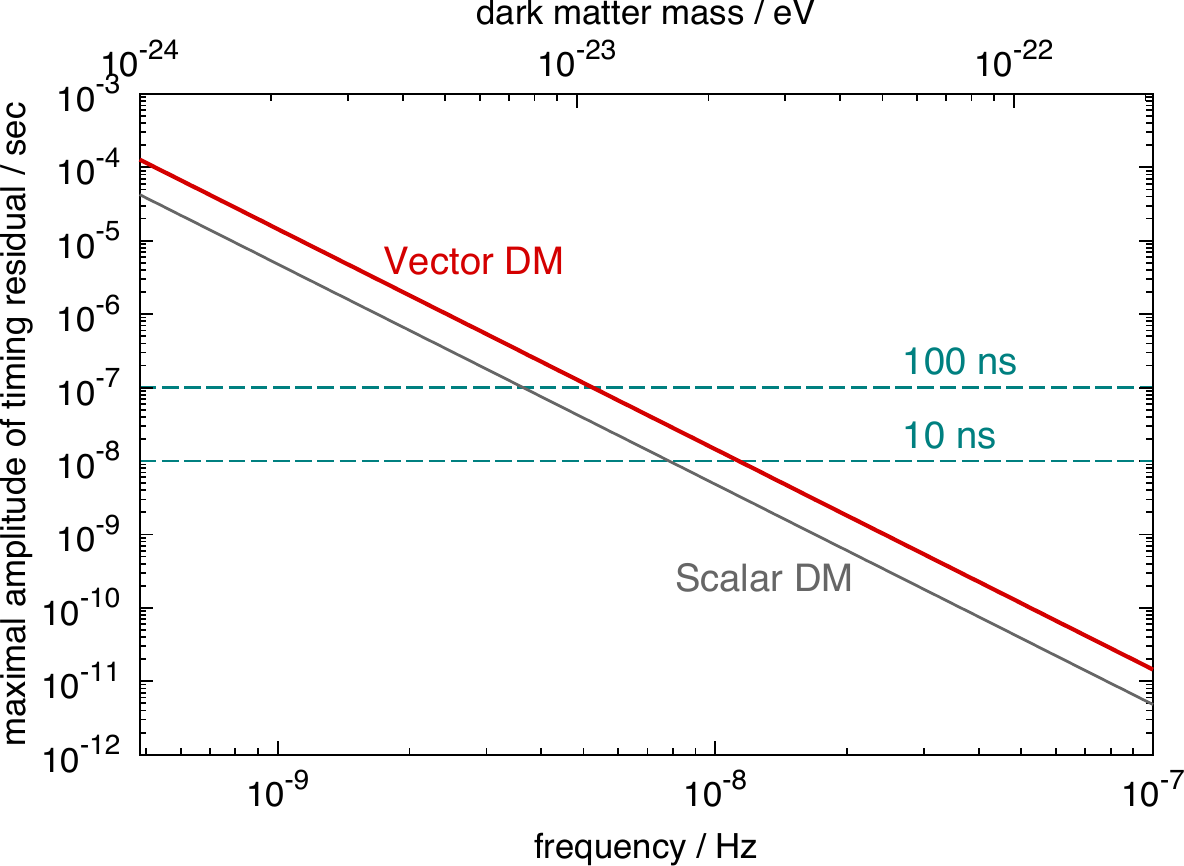}
  \caption{The maximum value of the amplitude of the pulsar timing residual due to the vector DM is indicated by the red line.
  The gray line corresponds to that of the scalar DM.
  Here we took the energy density of the DM as $\rho = 0.4 \, \mathrm{GeV}/\mathrm{cm}^{3}$.
  }
  \label{fig.VectorDM2}
\end{figure}

Finally, we assess the detectability of the vector DM by means of PTAs.
The maximal amplitude of the timing residual due to the vector DM oscillation is given by
\begin{align}
&\mathrm{max}|R_{\Psi} + R_{h}|
	\notag\\
&\quad= \mathrm{max}\left| \frac{\Psi_{osc}}{m} - \frac{1}{4} (1 + 3 \cos 2\theta) \frac{h_{osc}}{m} \right|
\notag \\
&\quad=  \mathrm{max}\left| - \frac{\pi G \rho}{3 m^{3}} - \frac{1}{4} (1 + 3 \cos 2\theta) \frac{8 \pi G \rho}{3m^{3}} \right|
\notag\\
&\quad= \frac{3 \pi G \rho}{m^{3}}
\notag \\
&\quad= 1.3 \times 10^{-7} \, \mathrm{sec}
	\left( \frac{\rho}{0.4 \, \mathrm{GeV}/\mathrm{cm}^{3}} \right)
	\left( \frac{10^{-23} \, \mathrm{eV}}{m} \right)^{3}.
\label{eq3.7.3}
\end{align}
In Fig. \ref{fig.VectorDM2}, the amplitudes of the timing residual estimated above are shown together with observational thresholds.
The threshold line at $100 \, \mathrm{ns}$ comes from
the best timing precision on the existing PTAs. For example,
 PSR J0437-4715 has a weighted root-mean-square of the timing residual
  $0.11 \, \mu\mathrm{s}$ \cite{Perera2019}.
In the near future, the Square Kilometre Array (SKA) project with suitable 250 pulsars may be able to measure with an accuracy of the order of $10 \, \mathrm{ns}$ \cite{Lommen2015}.

\section{Discussion and Conclusion}
\label{sec.conclusion}

We have studied the pulsar timing signal of the ultralight vector DM.
The vector DM in a galactic halo 
oscillates coherently and monochromatically with a specific frequency determined by its mass~\eqref{eq1.15}.
  The oscillation induces the time-dependent metric perturbations.
The metric perturbations yield a redshift of propagating electromagnetic fields.
Since the frequency of perturbations is typically in the nHz range,
 the redshift due to the DM is detectable by PTA experiments.
 The signal is monochromatic unlike the stochastic gravitational wave background e.g. 
 generated from the very early universe~\cite{Grishchuk:1974ny,Starobinsky:1979ty,Ito:2016fqp},
 the population of massive black hole 
 binaries~\cite{Ravi:2014nua,Sesana:2016yky,Kelley2017,Sesana:2017lnk} and
 the cosmic string network~\cite{Sanidas2012}.
Remarkably, we have shown that the pulsar timing residual due to the vector DM has a nontrivial angular dependence.
Especially, when the direction of the vector oscillation and the line of sight to the pulsar are parallel, the magnitude of the signal becomes maximum.
If the direction dependence of the pulsar timing residual is found, 
 it would be an evidence that the dominant component of
  the galactic DM is a vector field.
Fig. \ref{fig.VectorDM2} suggests that when the DM halo is dominated by the vector DM, that is, $\rho = 0.4 \, \mathrm{GeV}/\mathrm{cm}^{3}$, and its mass $m \lesssim 2 \times 10^{-23} \, \mathrm{eV}$, the amplitude of the 
 timing residual reaches $10 \, \mathrm{ns}$ at the maximum, 
which is the expected precision of the SKA project.
It is intriguing to observe the correlation between the statistical anisotropy~\cite{Watanabe:2009ct,Dulaney:2010sq,Gumrukcuoglu:2010yc,Watanabe:2010fh} of isocurvature perturbations induced by the vector~\cite{Nakayama2019} and the preferred direction detected by the PTAs. 

We assumed the vector DM has no couplings to the Standard Model.
Then, the constraints on the mass of the vector DM 
come from the superradiance of astrophysical black holes  
and the structure formation.
Since the superradiance reduces the rotation of a black hole
when the Compton length of the vector DM has the same order of
the gravitational radius of the black hole,
the existence of rotating stellar-mass black holes 
 excludes vector particles with masses, $5 \times 10^{-14}$--$2 \times 10^{-11} \, \mathrm{eV}$.
Also, vectors with lighter masses, $6 \times 10^{-20}$--$2 \times 10^{-17} \, \mathrm{eV} $, 
have been excluded from measurements of supermassive black holes, although they have less reliability \cite{Baryakhtar2017}. 
We should also keep in mind that the superradiance constraints are given
under the assumption that the self-interaction is sufficiently weak~\cite{Agrawal2018}.
Next, since the ultralight DM would suppress the structure formation on small scales, CMB data 
constrained masses of the axion-like ultralight scalar DM in the range $10^{-33} \leq m \leq 10^{-24} \, \mathrm{eV}$ \cite{Hlozek2017}.
Recently, the Lyman-$\alpha$ power spectrum in the ultralight scalar DM model was calculated using hydrodynamical simulations and compared with the observed data. 
It gave a lower limit on the 
mass of the scalar DM as $m \gtrsim 10^{-21} \, \mathrm{eV}$ \cite{Irsic2017, Kobayashi2017}.
Thus, the masses detectable by future PTAs are in tension with the above constraints
at least in the case of the scalar DM.
 Whether those can be directly applicable to the vector DM 
  is an issue to be considered.
 In any case, it is true that the PTA experiments can independently 
 constrain the energy density of the ultralight DMs
 and determine whether the dominant DM is vector or not.

As a future work, it is interesting to consider the detectability of the vector DM
with gravitational wave interferometers~\cite{Aoki:2016kwl}.
Our discussion would be applicable to the astrometric effects in a similar manner~\cite{Book2010}.
Moreover, it is important to evaluate gravitational waves from the vector DM during cosmological evolution~\cite{Soda:2017dsu,Kitajima:2018zco}.
We have extended the analysis of the scalar DM to the vector DM. In this line of thought, it is intriguing to study ultralight higher spin fields as the DM~\cite{Aoki:2016zgp,Aoki:2017cnz,Aoki:2017ixz}. 


\begin{acknowledgments}
The authors thank R. Kato for useful discussions.
A.\,I. was supported by Grant-in-Aid for JSPS Research Fellow and JSPS KAKENHI Grant No.JP17J00216.
J.\,S. was in part supported by JSPS KAKENHI
Grant Numbers JP17H02894, JP17K18778, JP15H05895, JP17H06359, JP18H04589.
J.\,S. was also supported by JSPS Bilateral Joint Research
Projects (JSPS-NRF collaboration) ``String Axion Cosmology.''
Discussions during the YITP workshop YITP-T-19-02 on ``Resonant instabilities in cosmology'' were useful for this work.
\end{acknowledgments}

\appendix

\section{Photon redshift due to perturbations on a flat spacetime}
\label{sec.photonredshift}

In this appendix, we calculate the photon redshift due to metric perturbations on a flat spacetime.
This section is based on~\cite{Book2010}, 
where the effect of gravitational waves on a photon trajectory is focused on.

Let the spatial origin $\mathbf{0}$ be the position of the observer.
The photon is observed by the observer at the time $t_{0}$.
We write the unperturbed world line of a photon traveling from a source to an observer as
\begin{align}
x_{0}^{\mu}(\lambda) = \omega_{0}(\lambda, - \lambda \mathbf{n}) + (t_{0}, \mathbf{0}),
\label{eqA.1.3}
\end{align}
where $\omega_{0}$ is a frequency, $\lambda$ is an affine parameter, 
and $\mathbf{n}$ is a unit vector from the observer to the source 
(so that $- \mathbf{n}$ is the propagation direction of the photon).
The trajectory \eqref{eqA.1.3} is chosen so that the photon reaches $x_{0}^{\mu} = (t_{0}, \mathbf{0})$ 
at $\lambda = \lambda_{obs} = 0$.
The unperturbed four-momentum of the photon is
\begin{align}
k_{0}^{\mu} = \frac{dx_{0}^{\mu}(\lambda)}{d\lambda}
=  \omega_{0}(1, -\mathbf{n}).
\label{eqA.1.4}
\end{align}
If the distance between the source and the observer is $|\mathbf{x}_{s}|$ , the affine parameter value $\lambda_{s}$ at which the photon is emitted by the source is given by
\begin{align}
\lambda_{s} = - \frac{|\mathbf{x}_{s}|}{\omega_{0}}.
\label{eqA.1.4.1}
\end{align}

Let us find the expression for the photon redshift up to the first order of perturbations.
We write the world line of a photon as a sum of an unperturbed part and  a perturbed part as
\begin{align}
x^{\mu}(\lambda) = x_{0}^{\mu}(\lambda) + x_{1}^{\mu}(\lambda)\ .
\label{eqA.1.5}
\end{align}
Similarly, the four-momentum is written as
\begin{align}
k^{\mu}(\lambda) = k_{0}^{\mu}(\lambda) + k_{1}^{\mu}(\lambda)\ .
\label{eqA.1.6}
\end{align}
We will consider the contributions from the scalar perturbations
  and the traceless part of the metric, separately.

\subsection{Scalar perturbations} \label{aps}

In the Newtonian gauge, the line element with scalar perturbations
 is written as
\begin{align}
ds^{2} = - (1 + 2\Phi) dt^{2} + (1 + 2\Psi) \delta_{ij} dx^{i} dx^{j} \ .
\label{eqA.1.1}
\end{align}
The scalar perturbations
 $\Phi$ and $\Psi$ are induced by the trace part of the
 energy-momentum tensor. 
Linearized Christoffel symbols are calculated as
\begin{align}
\Gamma^{0}_{00} = \dot{\Phi} \ , \quad
\Gamma^{0}_{i0} = \Gamma^{0}_{0i} = \partial_{i}\Phi \ , \quad
\Gamma^{0}_{ij} = \delta_{ij} \dot{\Psi} \ .
\label{eqA.1.2}
\end{align}

We use the photon geodesic equation
\begin{align}
\frac{dk^{\mu}}{d\lambda} = - \Gamma^{\mu}_{\nu \rho} k^{\nu} k^{\rho}.
\label{eqA.1.7}
\end{align}
From \eqref{eqA.1.4}$, k_{0}^{\mu}$ is independent of $\lambda$, so the left-hand side reduces to $dk_{1}^{\mu}/d\lambda$.
On the other hand, since Christoffel symbols are already the first order of the perturbations, only $k_{0}^{\mu}$ appears in the right-hand side.
Considering $\mu = 0$ component, we have
\begin{align}
\frac{dk_{1}^{0}}{d\lambda}
&= - \Gamma^{0}_{\nu \rho} k_{0}^{\nu} k_{0}^{\rho} \notag \\
&= - \Gamma^{0}_{00} k_{0}^{0} k_{0}^{0} - 2 \Gamma^{0}_{0i} k^{0}_{0} k^{i}_{0} - \Gamma^{0}_{ij} k^{i}_{0} k^{j}_{0} \notag \\
&= - \dot{\Phi} \omega_{0}^{2} - 2 \, \partial_{i}\Phi \,\omega_{0}^{2} (- n^{i}) - \dot{\Psi} \delta_{ij} \omega_{0}^{2} n^{i} n^{j} \notag \\
&= - \dot{\Phi} \omega_{0}^{2} + 2 \, \partial_{i}\Phi \,\omega_{0}^{2} n^{i} - \dot{\Psi} \omega_{0}^{2} \,.
\label{eqA.1.8}
\end{align}
In the last line, we used $\delta_{ij} n^{i} n^{j} = 1$.
We integrate \eqref{eqA.1.8} to obtain a four-momentum:
\begin{align}
k_{1}^{0}(\lambda) = \omega_{0}^{2} \int_{0}^{\lambda} d\lambda' \, 
(2 \partial_{i}\Phi \, n^{i}
- \dot{\Phi} - \dot{\Psi}
) + C,
\label{eqA.1.9}
\end{align}
where $C$ is a constant of integration.
We determine $C$ by considering the initial condition that the source emits a photon of the frequency $\omega_{0}$ at the affine parameter value $\lambda_{s}$.
In terms of the four-velocity of the source 
$u_{s}^{\mu}$, the condition is expressed as
\begin{align}
\omega_{0} = - g_{\mu\nu}(x_{s}) k^{\mu}(\lambda_{s}) u_{s}^{\nu}(\lambda_{s}).
\label{eqA.1.10}
\end{align}
Note that in the rest frame of the source, we obtain
\begin{align}
u^{0}_{s} = (1 + 2\Phi)^{-1/2} \simeq 1 - \Phi,
\label{eqA.1.11}
\end{align}
to the first order of perturbations.
Thus, from eq. \eqref{eqA.1.10}, we have
\begin{align}
\omega_{0} &= - g_{00}(x_{s}) k^{0}(\lambda_{s}) u^{0}_{s}(\lambda_{s}) \notag \\
&= \omega_{0} + \omega_{0}^{2} \int_{0}^{\lambda_{s}} d\lambda' \, (2 \partial_{i}\Phi \, n^{i} - \dot{\Phi} - \dot{\Psi} ) + C + \omega_{0} \Phi(x_{s}).
\label{eqA.1.11.1}
\end{align}
As a result, the constant $C$ must be
\begin{align}
C = - \omega_{0}^{2} \int_{0}^{\lambda_{s}} d\lambda' \, (2 \partial_{i}\Phi \, n^{i} - \dot{\Phi} - \dot{\Psi} )
 - \omega_{0} \Phi(x_{s}).
\label{eqA.1.12}
\end{align}

The photon frequency measured by the observer with the four-velocity
$u_{obs}^{\mu}$
can be found as 
$\omega_{obs} = - g_{\mu\nu}(x_{obs}) k^{\mu}(\lambda_{obs}) u_{obs}^{\nu}(\lambda_{obs})$, where
$x_{obs}^{\mu} = x^{\mu}(\lambda_{obs})$.
Using the expression $u_{obs}^{\mu} = (1 - \Phi(x_{obs}), 0, 0, 0)$ in the rest frame of the observer, we obtain
\begin{align}
\omega_{obs} &= - g_{00}(x_{obs}) k^{0}(\lambda_{obs}) u_{obs}^{0}(\lambda_{obs}) \notag \\
&=  \omega_0 + \omega_{0}^{2} \int_{\lambda_{s}}^{0} d\lambda' \, (2 \partial_{i}\Phi \, n^{i} - \dot{\Phi} - \dot{\Psi} )
\notag \\
&\quad- \omega_{0} \Phi(x_{s}) + \omega_{0} \Phi(x_{obs}) \,.
\label{eqA.1.13}
\end{align}
We can use the relation $t = \omega_{0} \lambda + t_{0}$ along the unperturbed photon geodesic,  $t = t_{0}$ at $\lambda = 0$, and 
$t = t_{0} - |\mathbf{x}_{s}|$ at $\lambda = \lambda_{s}$ to rewrite this equation.
In addition, the total derivative with respect to $\lambda$ is given by $d/d\lambda = \omega_{0} \partial_{t} - \omega_{0} n^{i} \partial_{i}$.
Then, eq.\,\eqref{eqA.1.13} can be rewritten as
\begin{align}
\omega_{obs} 
&= \omega_0 + \omega_{0} \int_{t_{0} - |\mathbf{x}_{s}|}^{t_{0}} \! dt' \, 
[ \partial_{i}\Phi(t') \, n^{i} - \partial_{i}\Psi(t') \, n^{i} ] 
\notag \\
&\quad- \omega_{0} \Psi(x_{obs}) + \omega_{0} \Psi(x_{s}).
\label{eqA.1.13.1}
\end{align}
We now estimate the magnitude of the signal in PTA measurements.
The distance to a pulsar is typically $|\mathbf{x}_{s}| \gtrsim 100 \, \mathrm{pc}$ \cite{Perera2019}, which is much longer than 
 $m^{-1} = 0.6 \, \mathrm{pc} \times (10^{-23}\,\mathrm{eV}/m)$.
Therefore, the integrand of the second term in \eqref{eqA.1.13.1} is rapidly oscillating, and hence becomes small after the integration.
Moreover, since the spatial derivative gives  $k = 2\pi / \lambda_{\mathrm{dB}}$ and $m^{-1}$ is factored out from the integral,
 the second term in \eqref{eqA.1.13.1} is suppressed by a factor $k/m = v \sim 10^{-3}$ compared to the last two terms (see the discussion in Section \ref{sec.metricperturbation}).
 Thus, we can neglect the second term.
As a result, the redshift  of the photon is given by
\begin{align}
z = \frac{\omega_{0} - \omega_{obs}}{\omega_{0}} &=  \Psi(x_{obs}) - \Psi(x_{s}) \notag \\
&=\Psi(t_{0}, \mathbf{0}) - \Psi(t_{0}-|\mathbf{x}_{s}|, \mathbf{x}_{s}) 
\,.
\label{eqA.1.14}
\end{align}

\subsection{Traceless part of metric perturbations} \label{apt}

We apply the above discussion to the traceless part of  metric perturbations.
We write the line element as
\begin{align}
ds^{2} = - dt^{2} + (\delta_{ij} + h_{ij}) dx^{i} dx^{j},
\label{eqA.2.1}
\end{align}
where $h_{ij}$ corresponds to the traceless perturbations.
From this metric, linearized Christoffel symbols are calculated as
\begin{align}
\Gamma^{0}_{ij} &= \frac{1}{2} \dot{h}_{ij}, \notag \\
\Gamma^{i}_{0j} &= \Gamma^{i}_{j0} = \frac{1}{2} {{\dot{h}}^{i}}_{~j}, \notag \\
\Gamma^{i}_{jk} &= \frac{1}{2} (\partial_{k}{h^{i}}_{j} + \partial_{j}{h^{i}}_{k} - \partial^{i}{h_{jk}}).
\label{eqA.2.2}
\end{align}
In this case, the timelike component of the four-momentum of the photon obeys
\begin{align}
\frac{dk_{1}^{0}}{d\lambda} = - \Gamma^{0}_{ij} k_{0}^{i} k_{0}^{j}
= - \frac{1}{2} \dot{h}_{ij} \omega_{0}^{2} n^{i} n^{j}.
\label{eqA.2.3}
\end{align}
Integrating this expression yields
\begin{align}
k_{1}^{0}(\lambda) = - \frac{1}{2} \omega_{0}^{2} n^{i} n^{j} \int_{0}^{\lambda} d\lambda' \, \dot{h}_{ij}(\lambda') + C',
\label{eqA.2.4}
\end{align}
where $C'$ is a constant of integration.
We use the condition \eqref{eqA.1.10} to determine $C'$.
In the rest frame of the source, its four-velocity is $u_{s}^{\nu} = (1, 0, 0, 0)$.
So we have
\begin{align}
\omega_{0} &= - g_{\mu\nu}(x_{s}) k^{\mu}(\lambda_{s}) u_{s}^{\nu}(\lambda_{s}) \notag \\
&= \omega_{0} - \frac{1}{2} \omega_{0}^{2} n^{i} n^{j} \int_{0}^{\lambda_{s}} d\lambda' \, \dot{h}_{ij}(\lambda') + C' \ .
\label{eqA.2.4.1}
\end{align}
Thus, 
\begin{align}
C' = \frac{1}{2} \omega_{0}^{2} n^{i} n^{j} \int_{0}^{\lambda_{s}} d\lambda' \, \dot{h}_{ij}(\lambda').
\label{eqA.2.5}
\end{align}
The frequency of a photon measured by an observer is calculated as
 $\omega_{obs} = - g_{\mu\nu}(x_{obs}) k^{\mu}(\lambda_{obs}) u_{obs}^{\nu}(\lambda_{obs})$.
Since $u_{obs}^{\mu} = (1, 0, 0, 0)$ in the rest frame of the observer, 
we find
\begin{align}
\omega_{obs} &= - g_{\mu\nu}(x_{obs}) k^{\mu}(\lambda_{obs}) u_{obs}^{\nu}(\lambda_{obs}) \notag \\
&= \omega_{0} + \frac{1}{2} \omega_{0}^{2} n^{i} n^{j} \int_{0}^{\lambda_{s}} d\lambda' \, \dot{h}_{ij}(\lambda').
\label{eqA.2.6}
\end{align}
To the first order of perturbations, $h_{ij}(\lambda)$ means $h_{ij}(x_{0}(\lambda)) = h_{ij}(\omega_{0} \lambda + t_{0}, - \omega_{0} \lambda \mathbf{n})$.
Hence, we obtain
\begin{align}
\frac{dh_{ij}(\lambda)}{d\lambda} &= \dot{h}_{ij} \frac{d}{d \lambda}(\omega_{0} \lambda + t_{0}) + \partial_{k} h_{ij} \frac{d}{d \lambda} (-\omega_{0} \lambda n^{k}) \notag \\
&= \omega_{0} \dot{h}_{ij} - \omega_{0} n^{k} \partial_{k} h_{ij}.
\label{eqA.2.7}
\end{align}
Using this relation, we can rewrite \eqref{eqA.2.6} as
\begin{align}
&\omega_{obs}
\notag \\
&= \omega_{0} + \frac{1}{2} \omega_{0}^{2} n^{i} n^{j} \int_{0}^{\lambda_{s}} d\lambda' \, \dot{h}_{ij}(\lambda') \notag \\
&= \omega_{0} + \frac{1}{2} \omega_{0} n^{i} n^{j} \int_{0}^{\lambda_{s}} d\lambda' \, \left( \frac{dh_{ij}(\lambda')}{d\lambda'}  + \omega_{0} n^{k} \partial_{k}h_{ij}(\lambda') \right) \notag \\
&= \omega_{0} + \frac{1}{2} \omega_{0} n^{i} n^{j} \left[ h_{ij}(t_{0} - |\mathbf{x}_{s}|, \mathbf{x}_{s}) - h_{ij}(t_{0}, \mathbf{0}) \right] 
\notag \\
&\quad+ \frac{1}{2} \omega_{0} n^{i} n^{j} \int_{t_{0}}^{t_{0} - |\mathbf{x}_{s}|} dt' \, n^{k} \partial_{k}h_{ij}(t', \mathbf{x}(t')).
\label{eqA.2.8}
\end{align}
The third term is again negligible with the same argument as the case for the scalar perturbations.
Thus, we have the expression for the redshift as
\begin{align}
z = \frac{\omega_{0} - \omega_{obs}}{\omega_{0}} &= \frac{1}{2} n^{i} n^{j} \left[  h_{ij}(t_{0}, \mathbf{0}) - h_{ij}(t_{0} - |\mathbf{x}_{s}|, \mathbf{x}_{s}) \right] \notag \\
&= \frac{1}{2} n^{i} n^{j} \left[  h_{ij}(x_{obs}) - h_{ij}(x_{s}) \right].
\label{eqA.2.9}
\end{align}
In contrast to \eqref{eqA.1.14}, the redshift 
depends on the direction to the source $\mathbf{n}$.

Finally, we note that our discussion can be applied to 
periodic pulses radiated from a pulsar and then 
$\omega_{0}$ and $\omega_{obs}$ in eqs. (\ref{eqA.1.14}) and (\ref{eqA.2.9}) 
are identified as the angular frequencies of the pulses at the pulsar and the observer, respectively.

\bibliography{biblio_2}

\end{document}